\documentclass[fleqn,10pt]{wlscirep}
\usepackage[utf8]{inputenc}
\usepackage[T1]{fontenc}

\usepackage{graphicx}
\usepackage{dcolumn}
\usepackage{bm}
\usepackage{color}
\usepackage{amsmath}
\usepackage{wasysym}
\usepackage{physics}
\usepackage{eucal}
\usepackage{CJK}

\usepackage{amssymb}

\definecolor{drkgreen}{rgb}{0.0, 0.5, 0.0}
\definecolor{violet}{rgb}{0.5, 0.0, 1.0}

\title{Braiding properties of worldline configurations in hardcore lattice bosons }

\author[1,2,3,*,+]{Fabio Lingua}
\author[2,+]{Wei Wang (王巍)}
\author[1]{Liana Shpani}
\author[1]{Barbara Capogrosso-Sansone}
\affil[1]{Department of Physics, Clark University, Worcester, Massachusetts 01610, USA}
\affil[2]{Tsung-Dao Lee Institute, Shanghai Jiao Tong University, Shanghai 201210, China}
\affil[3]{Department of Physics and Astronomy, Dartmouth College, Hanover, New Hampshire 03755, USA}
\affil[*]{lingua.fabio@gmail.com}

\affil[+]{these authors contributed equally to this work}

\date{\today}

\begin{abstract}
In this manuscript, we study braiding properties of worldline configurations for a variety of ground-states of hardcore Bose-Hubbard models in two dimensions. Configurations are collections of particle paths and result from the path-integral formulation of statistical mechanics. For hard-core bosons, configurations can be seen as geometric braids and therefore can be assigned a certain topological structure, i.e. a way to classify braiding events among worldlines.
By means of Monte Carlo calculations, we study superfluid phase and a variety of insulating phases and observe that ground-states of different quantum phases correspond to different probability distributions of braiding properties.
\end{abstract}

\begin{document}
\begin{CJK*}{UTF8}{gbsn} 

\flushbottom
\maketitle
%
%
\thispagestyle{empty}

\section*{Introduction}\label{intro}
In 1948, Feynman introduced the path-integral approach~\cite{Feynman1948} of quantum mechanics.
Within this approach, transition probability amplitudes
can be computed in terms of superpositions of classical paths so that all possible routes are taken into account but only those which superpose constructively contribute the most to the quantum physical process.
Therefore, there exists a parallel between a certain quantum process and a {\em set} of classical possibilities which result in the same outcome. A similar parallel exists within the Feynman's path-integral formulation of statistical mechanics where one works in imaginary time (inverse temperature)\cite{FeynmanBook}. Within this framework, each quantum particle of a many-body system is mapped onto a trajectory (also known as worldline) in space and imaginary-time, so that the quantum system is now described in terms of collections of worldlines in space and imaginary-time, i.e worldline configurations.  
Therefore, there exists a correspondence between the D-dimensional quantum system and a (D+1)-dimensional classical system~\cite{RevModPhys.69.315}.

 In this work, we use this correspondence
and consider the possibility of describing properties of quantum strongly-correlated many-body ground-states in terms of properties of (constructively superposing) classical `paths'.  The relation between the set of higher dimensional classical objects and the quantum system is well-understood in terms mathematical averages, e.g. correlation functions, expectation values of physical observables. Nonetheless, one might wonder if, by its very nature, the path-integral description can provide further insight on properties of ground-states. Can a classification of worldline configurations in terms of their mathematical properties highlight system properties beyond what one learns from computation of expectation values? To clarify what we mean, let us consider the fact that the kinetic-energy term in the Hamiltonian drives worldlines to intertwine with each others.  For hard-core lattice bosons, these intertwined set of worldlines can be interpreted as braids and therefore can be assigned a topological structure, i.e. a way to classify braiding events among worldlines. In this respect, configurations can be considered as a visualization of the interplay between the kinetic and interaction terms of the Hamiltonian. The interplay between these two terms is ultimately responsible for ground-state properties. Therefore, one can expect that braiding properties of configurations should, at the very least, reflect already known ground-state properties, and likely
being useful to investigate new aspects of quantum phases.

Here, we begin to test this idea by showing that a simple topological invariant associated to the topological structure of worldline configurations can be used to differentiate ground-states of hardcore Bose-Hubbard models. More specifically, by means of large-scale Monte Carlo simulations, we collect a statistically relevant sample of worldline configurations for a variety of ground-states such as checkerboard, stripe and valence-bond solids, and  $\mathbb{Z}_2$ topologically ordered phases, and differentiate them by studying the statistics of a simple topological invariant associated to the braiding structure of configurations.

\section{Review of Path-integral formalism}
The imaginary-time path-integral formulation of the density matrix is the foundation of path integral Monte Carlo (PIMC) algorithms~\cite{Ceperley1995,Prokofev1998}. Here, we consider lattice hard-core bosons in two dimensions. Let us define $\ket{\alpha}=\ket{0,1,\cdots}$, where 1 corresponds to occupation of a hardcore boson and 0 to no occupation in the Fock basis. The generic Bose-Hubbard-type model describing the system has the form:
\begin{equation}
\label{eq2} H=-\sum_{\langle i,j\rangle } (t_{ij}a_i^\dagger a_j+\mathrm{H.c.})+H_0 -\mu\hat N
\end{equation}
where the first term is the hopping between nearest neighboring sites $i$ and $j$ with strength $t_{ij}$ (in the following denoted by $H_1$), while the second term is an arbitrary diagonal term, e.g., $\sum_{(i,j)} V_{ij}a_i^\dagger a_i a_j^\dagger a_j$. $\hat N$ is particle number operator, and $\mu$ is the chemical potential which we use in our simulations to control the number of particles. In the following, we consider $t_{ij}>0$.

Within the path-integral formalism~\cite{Ceperley1984,Ceperley1995,multiworm}, the partition function at temperature $k_BT=1/\beta$
\begin{equation}
\label{eq3} \mathcal{Z}_\beta= \Tr \mathrm{e}^{-\beta H}= \sum_\alpha \mel{\alpha}{\mathrm{e}^{-\beta H}}{\alpha}=\int_{\phi\in\mathcal{C}} ~\omega_\phi \; ,
\end{equation}
is an integral of weights $\omega_\phi>0$ of worldline configurations (see Eq.~\ref{weights} for more details), where $\phi$ is a combination of continuous and discrete indexes which uniquely specifies the configuration, and $\mathcal{C}$ is the set of all configurations. A configuration $\phi$ is a collection of worldlines (see Fig. 1) where each worldline represents the path of a particle in imaginary-time and space. In the interaction picture, $\phi$ is specified by a sequence of imaginary time instants at which single hopping events happen $0<\tau_1<\ldots<\tau_{n-1}<\beta$, and the corresponding sequence of Fock states $\{\ket{\alpha},\ket{\alpha_1},\ldots,\ket{\alpha_{n-1}},\ket{\alpha}\}$, where state $\ket{\alpha_{i}}$ is the state at time $\tau_i$.  Note that, by definition of $\mathcal{Z}_\beta$, periodic boundary conditions in imaginary time hold, that is, the configuration starts and ends with the same occupation of lattice sites specified by a Fock state $\ket{\alpha}$. Within this framework, weights $\omega_\phi$ can be expressed as~\cite{Ceperley1984,Ceperley1995,multiworm}:
\begin{equation}
\omega_\phi = (-1)^n\mathrm{e}^{-\beta E_\alpha}\mel{\alpha}{H_1(\tau_n)}{\alpha_{n-1}}\cdots\mel{\alpha_{1}}{H_1(\tau_{1})}{\alpha} \; ,
\label{weights}
\end{equation}
where $H_1(\tau)=\mathrm{e}^{H_0\tau}H_1\mathrm{e}^{-H_0\tau}$.

\section{Braiding properties of worldline configurations}\label{topo_str}

\begin{figure}[ht]
\centering
    \includegraphics[width=1\textwidth]{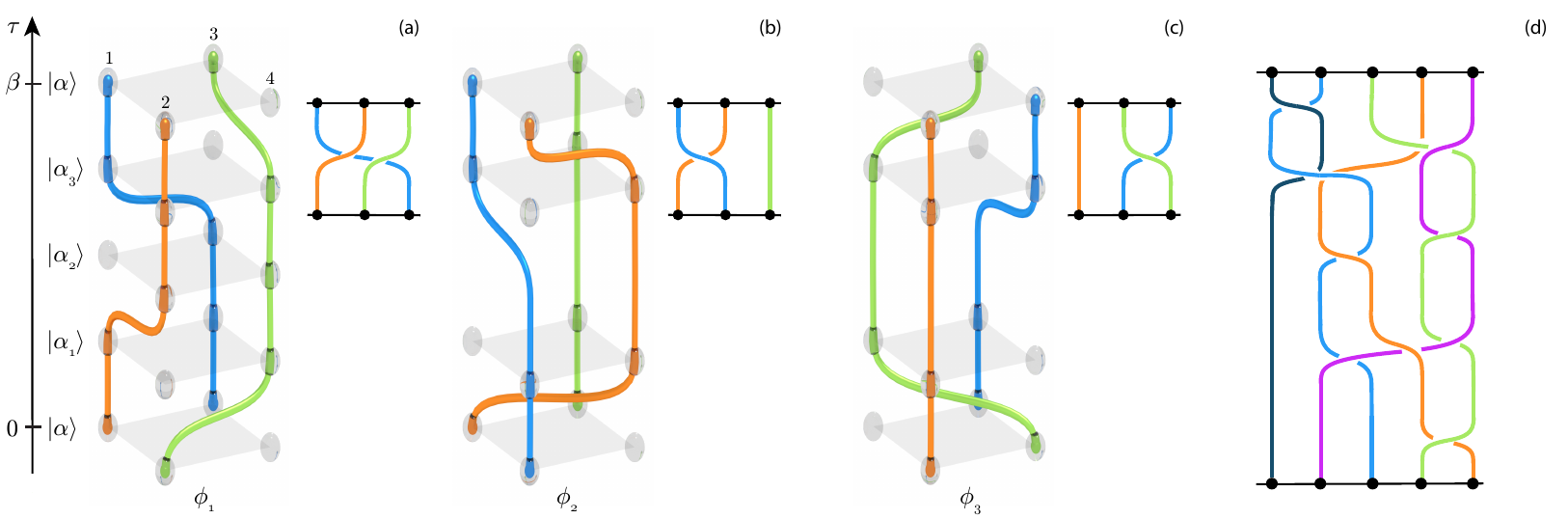}
    \caption{Panels (a), (b) and (c): three examples of configurations of 3 particles on a $2\times2$ lattice. Configurations are collections of particle-paths in space and imaginary-time $\tau$ and are specified by a sequence of Fock states $\ket{\alpha_i}$ ordered in $\tau$. Smaller figures are a 2D depiction of the geometric braid only including occupied sites and in which cut worldlines  are the ones behind another worldline. Particles are indistinguishable, colors are only meant to guide the eye. Panel (d): example of a more complex geometric braid involving 5 particles.}
    \label{fig:1}
\end{figure}

Let us consider a worldline configuration $\phi$  as specified by the sequence of imaginary time instants $0<\tau_1<\ldots<\tau_{n-1}<\beta$ and  corresponding Fock states $\{\ket{\alpha},\ket{\alpha_1},\ldots,\ket{\alpha_{n-1}},\ket{\alpha}\}$.
Expectation values  entering  $\omega_\phi$ are nonzero, i.e. $\mel{\alpha_{m+1}}{H_1(\tau_{m+1})}{\alpha_{m}}\ne0$, implying that $\ket{\alpha_m}$ and $\ket{\alpha_{m+1}}$ differ only on one pair of sites $(i,j)$, i.e. $\ket{\alpha_{m+1}}=a_i^\dagger a_j\ket{\alpha_m}$. This, along with the hard-core nature of the particles, implies that there is no intersection between any two worldlines. We can therefore conclude that a configuration $\phi$ can be uniquely identified as a {\em geometric braid}, i.e. $N$ non-crossing strands in the space $[0,\beta]\times\mathcal{S}$~\cite{Kassel2008}, and, as such, it corresponds to an element of the braid group $\mathcal{B}_N(\mathcal{S})$~\cite{Delima2018}, where $N$ is the particle number and $\mathcal{S}$ is the surface in which the lattice is embedded. Notice that an element in $\mathcal{B}_N(\mathcal{S})$ is an equivalent class of geometric braids. In other words,
two worldline configurations correspond to different braids or have different topological structures if it is impossible to continuously deform worldlines of one configuration into those of the other without cutting worldlines, otherwise they correspond to the same braid. Therefore, the topological structure of $\phi$ is specified by how particles move $around$ each other via sequential hopping $a_i^\dagger a_j$. A more rigorous definition of topological structure is given in~\ref{App_A}, where we define topological structure based on homotopy rather than isotopy. To fix the idea, in Fig.~\ref{fig:1}, we show simple examples of worldline configurations and associated braid diagrams (i.e. the geometric representations of the braid group elements). Notice that different configurations can feature the same topological structure (panels (b) and (c)). Fig.~\ref{fig:1} (d) shows how braids may easily become more complex as the number of particles is increased.

In this work, we investigate quantum systems in the limit of zero temperature. A basic fact derived from the path-integral formalism is that the expectation value of observables in the ground-state $\ket{\Psi}$ is given by the limit
\begin{equation}
\label{eq4}\expval{O}=\lim_{\beta\rightarrow\infty}\int_{\phi\in\mathcal{C}} ~ \frac{\omega_\phi O_\phi}{\mathcal{Z}_\beta}
\end{equation}
where $\omega_\phi/\mathcal{Z}_\beta$ is a probability density and $O_\phi$ is the value of $O$ in configuration $\phi$. The convergence of the above limit implies that the physics of ground-state can be captured with desired accuracy by the path-integral formalism  for some large enough $\beta$. That is,  once a large enough value of $\beta$ is reached, i.e. the system is effectively in the ground-state, system properties, as determined by the set of {\em{relevant}} configurations,  become $\beta$-independent. By relevant we mean the configurations that contribute the most, that is, the ones for which the product of the number of configurations with similar non-negligible weight times the weight is maximum. According to the correspondence between the D-dimensional quantum system and a (D+1)-dimensional classical system, properties of the quantum system possess counterparts in the classical one. Given that the topological structure can represent the complexity of correlated classical trajectories \cite{Thiffeault2010}, we expect that, for large enough values of $\beta$, the relevant topological structures would reflect certain ground-state properties (see examples below).

To elucidate this expectation, let us discuss what basic braiding properties of configurations one would expect for some well-known insulating ground-states (checkerboard solid, valence bond solid, and $\mathbb{Z}_2$ spin liquid). We will later discuss how our numerical results provide good evidence in this direction. Let us start with the consideration that according to Eq.~\ref{eq4} with $O$ replaced by $\dyad{\alpha}{\alpha}$
the Fock states entering the ground-state expansions are the same as the ones participating in the relevant configurations (see \ref{AppA0} for details).  Deep in the checkerboard solid, the ground-state expansion is dominated by $\ket{\alpha_{gs}}$~\cite{Hebert2001} which corresponds to one of the two classical checkerboard configurations (for large systems, the energy penalty to go from one to the other classical configuration is high).  Then, any relevant worldline configuration $\phi$ in the checkerboard phase is expected to be dominated by $\ket{\alpha_{gs}}$ (that is, the majority of the imaginary time between 0 and $\beta$ will be spent in $\ket{\alpha_{gs}}$) and be either a trivial braid or a braid with only few worldlines braided together locally. This ensures that Fock states in the sequence characterizing $\phi$ minimally differ from  $\ket{\alpha_{gs}}$, i.e. they have almost the same occupation of bosons as $\ket{\alpha_{gs}}$. Deep in the valence bond solid at filling 1/3 on Kagome lattice~\cite{IsakovPRL2006}, the ground-state expansion is dominated by Fock states corresponding to the classical hexagon-solid-backbone of holes with only isolated hexagons harboring local fluctuations of occupation numbers of bosons. Therefore, relevant $\phi$ may only have few worldlines braided together. These worldlines prevalently start and end on the hexagons where local resonances exist. Finally, deep in the $\mathbb{Z}_2$ spin liquid on Kagome lattice, the ground-state is an expansion that includes Fock states which can be dramatically different from each other in terms of occupation number of bosons. Nonetheless, every pair of states entering the expansion can be connected by sequential operations of the form $a_ia_j^\dagger$~\cite{Wang2017} with small energy cost.
Hence, deep in this phase, there can be relevant $\phi$ such that very different states (in terms of bosonic occupation number) appear in the sequence characterizing $\phi$. As a consequence, one may expect that, unlike the case of checkerboard and valence-bond solids (for which only worldines simply and locally braided with each other may appear in a configuration along with non-braided wordlines), in the case of the $\mathbb{Z}_2$ spin liquid, complex non-local braids connecting the states entering the ground-state expansion may appear in a configuration.

In the next Section, we discuss some properties of topological structures of relevant configurations.   To this end, we replace the $O_\phi$ in Eq.~\ref{eq4} by a
topological invariant (see below)  and observe that distinct relevant topological structures correspond to ground-states of different quantum phases.

\section{Numerical results}
The simplest topological invariant that can be used to {\em partially} characterize the topological structure of configurations is the statistics of permutation cycles \cite{Feynman1953,Ceperley1995}. Permutation cycles are obtained by gluing together worldlines at imaginary time $\tau=\beta$ and $\tau=0$. 
Their length only depends on the {\em minimal} number of braiding events occurring between worldlines participating in a cycle and does not detect braiding events among different permutation cycles. Hence, the length of permutation cycles does not provide a complete characterization of a topological structure. For example, a permutation cycle involving two worldlines can have any odd number (1,3,5,...) of braiding events between the two. Nonetheless, despite their simplicity, permutation cycles and their statistics are already able to distinguish ground-states of different quantum phases. To be clear, by statistics of permutation cycles, we do not simply mean their overall statistical weight (see e.g.~\cite{PhysRevLett.83.2687,PhysRevE.76.051109}). Rather, we characterize each configurations by the permutation cycles in it and calculate the probability associated to each combination of permutation cycles.

Formally, we use the topological invariant ${\vec q_\phi}=(n_1, n_2, ..., n_N)$, where $n_l$ is the number of permutation cycles of length $\lambda=l\times\beta$ appearing in the configuration, and $N$ is the total number of particles. For example, configurations in Fig.~\ref{fig:1} are characterized by $\vec q_{\phi_1}=(0,0,1)$ and $\vec q_{\phi_2}=\vec q_{\phi_3}=(1,1,0)$. By collecting statistics on $\vec q_\phi$, one can compute the probability that topological structures characterized by $\vec{q}_\phi$ appear:
\begin{equation}
p_\mathcal{\tilde T} = \int_{\phi\in\mathcal{C_{\tilde T}}} \frac{\omega_\phi}{\mathcal{Z}_\beta }\; ,
\end{equation}
where $\mathcal{C_{\tilde T}}$ is the set of all configurations characterized by the same topological invariant $\vec{q}_\phi$. These configurations do not necessarily have the same topological structure.  We calculate probability $p_\mathcal{\tilde T}$ for all measured $\vec{q}_\phi$ and
obtain a probability distribution that can be interpreted as a spectrum of topological structures capable of distinguishing different ground-states.

We perform quantum Monte Carlo simulations of the extended Bose Hubbard model: \begin{equation}
    H = -t\sum_{\langle ij\rangle}a^\dag_i a_j + H_0\label{Hsq}
\end{equation}
where $H_0$ is the diagonal part of $H$ in the Fock basis and $\langle ij\rangle$ refers to sum over nearest neighbors. Here, we choose hopping $t>0$. We consider four $H_0$: (1) $H_0=V\sum_{\langle ij\rangle}n_i n_j$ in a square lattice which, at filling factor 1/2, can stabilize a checkerboard (CB) phase~\cite{Hebert2001}; (2) $H_0=V\sum_{ij}\frac{n_i n_j}{r_{ij}^3}$ in a square lattice which can stabilize a stripe (STR) phase at filling factor $1/3$~\cite{Capogrosso2010} (here, $r_{ij}$ is the distance between site $i$ and $j$ and we set a cut-off to $r_{ij}=4$); (3) $H_0=V\sum_{\hexagon}n_i n_j$, where the sum over $\hexagon$ refers to the sum between sites on the same hexagon of the Kagome lattice, which, at filling factor 1/2 and 1/3 can stabilize a $\mathbb{Z}_2$ topologically ordered phase~\cite{Isakov2011,Roychowdhury2015}; (4) $H_0=V\sum_{\langle ij\rangle}n_i n_j$  on the Kagome lattice, which at $1/3$ stabilizes a valence-bond solid (VBS) \cite{IsakovPRL2006}. We remind to \ref{AppE} for details on the critical values of $V/t$ at which the considered transitions occur. For all models we consider periodic boundary conditions.
We measure topological invariant $\vec{q}_\phi$ for a number of configurations of the order of $10^4 $ to $10^5$. From these measurements, we compute $p_\mathcal{\tilde T}$.

\begin{figure}
    \centering
    \includegraphics[width=0.9\textwidth]{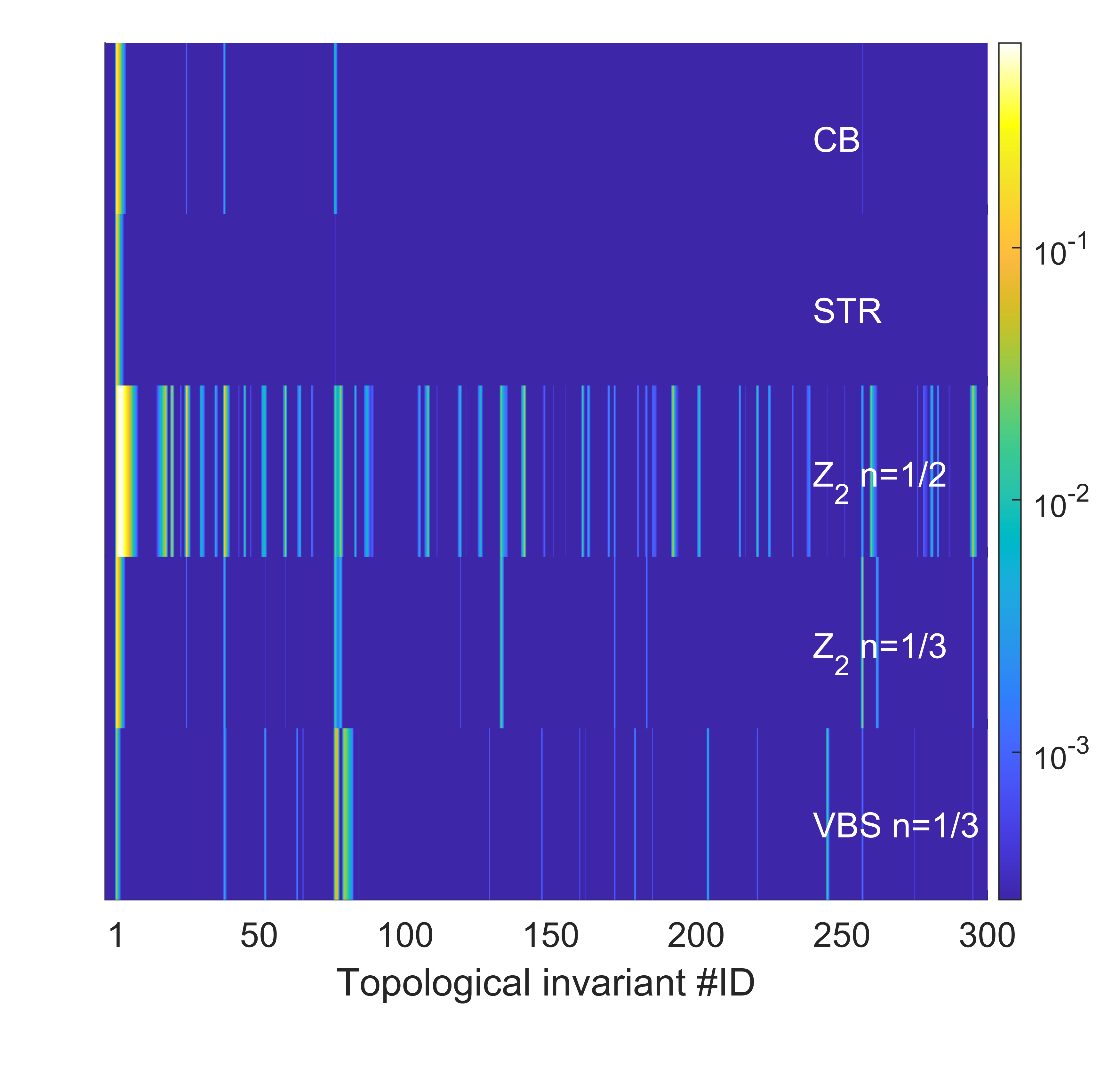}
    \caption{Distribution $p_\mathcal{\tilde T}$ of ground-states of different insulating quantum phases for comparable system-sizes (in terms of total number of sites): $L=10$ for square lattice and $L=6$ unit cells for Kagome lattice. From top to bottom: checkerboard (CB) at $V/t=20$, stripe solid (STR) at $V/t=60$, $\mathbb{Z}_2$ topologically ordered insulator ($Z_2$)  for $n=1/2$ and $n=1/3$ at $V/t=15$, and Valence bond Solid (VBS) at $V/t=15$. Inverse temperature $\beta t=18$.}
    \label{fig:topospectra}
\end{figure}

In Fig.~\ref{fig:topospectra}, we plot probability distribution $p_\mathcal{\tilde T}$ corresponding to each ground-state of the insulating phases as a color map. Color maps can be interpreted as topological spectra. These spectra manifest
visible differences for different ground-states.
Color maps 
are plotted as function of a label which orders all topological invariants found. We label them according to their average length of permutation cycles by excluding $1\beta$-long permutation cycles (see \ref{App_C} for details). Fig.~\ref{fig:topospectra} shows topological spectra of CB at $V/V_c=10$, STR at $V/V_c=4$, and $\mathbb{Z}_2$ and VBS at $V\approx2V_c$.
These ground-states are characterized by topological invariants that feature small permutation cycles, as expected for insulating phases, with the majority of worldlines in $1\beta$ permutation cycles and a few worldlines mainly participating to $2\beta$ and/or $3\beta$ permutation cycles. Longer permutation cycles  such as $4\beta$ and $5\beta$ only appear in ground-states of $\mathbb{Z}_2$ phases. 

\begin{figure*}
    \centering
    \includegraphics[width=0.9\textwidth]{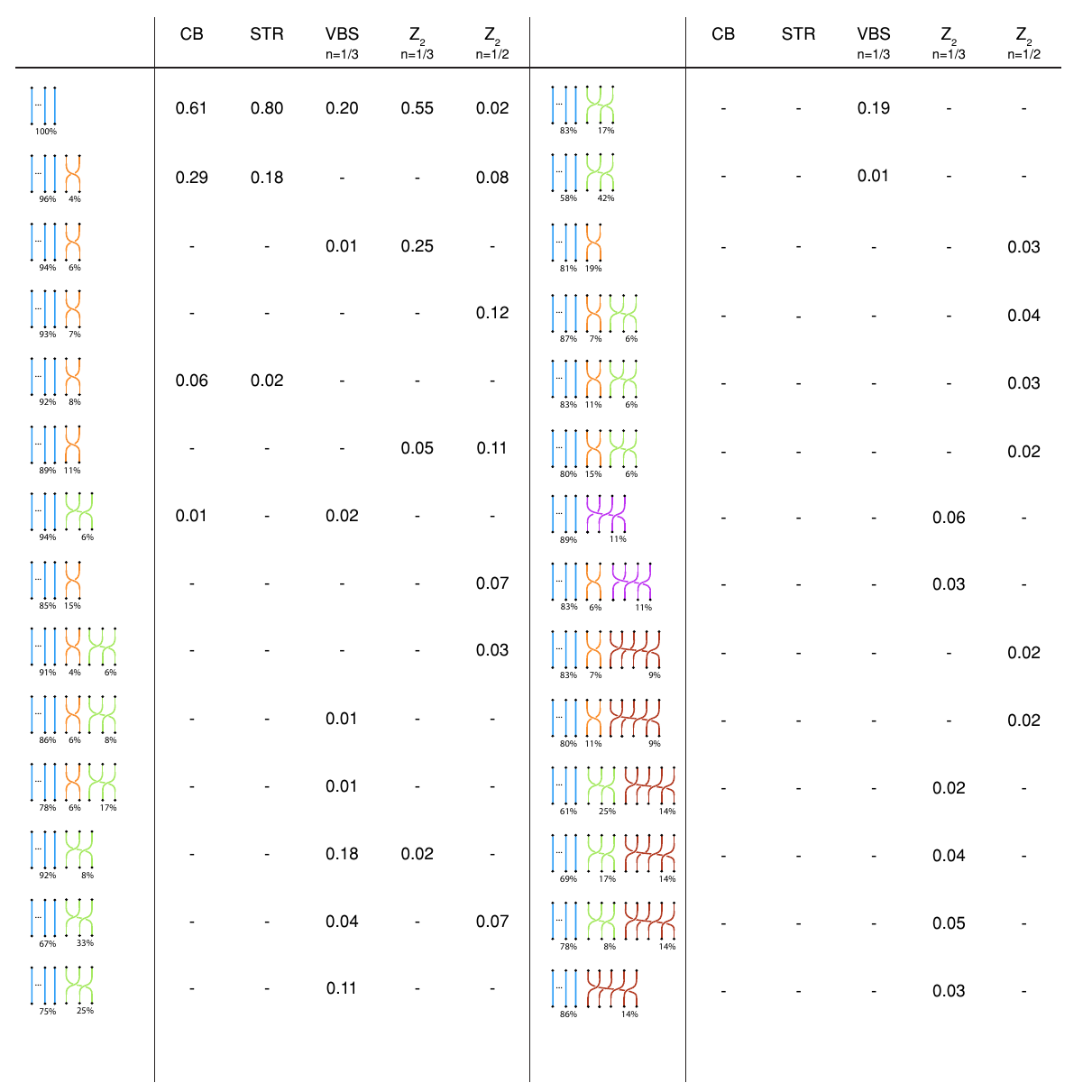}
    \caption{Arrangement of permutation cycles in ground-states of different insulating quantum phases. Numbers in the table represents the probability of each arrangement to appear in the related phase. The percentage in the diagrams refers to the components of $\vec{p}_\phi$ (i.e. the fraction of particles participating in permutation cycles of that length). CB for $L=10$ and $V/t=20$; STR for $L=12$ and $V/t=20$; VBS at $n=1/3$ for $L=6$ unit cells and $V/t=30$; $\mathbb{Z}_2$ at $n=1/3$ and $n=1/2$ for for $L=6$ unit cells and $V/t=15$. }
    \label{fig:topostruct}
\end{figure*}

In the table of Fig.~\ref{fig:topostruct} we sketch the most probable (with a probability greater than 0.01) arrangement of permutation cycles (first column), with corresponding probability  to appear in the expansion of the partition function (remaining columns).
Each arrangement of permutation cycles is represented in terms of the fraction of worldlines (percentages under the braid diagrams) participating to the corresponding permutation cycle: $1\beta$-long permutation cycles corresponding to straight worldlines (blue), $2\beta$-long permutation cycles corresponding to two worldlines braiding together (orange), $3\beta$-long permutation cycles and longer corresponding to three or more worldlines braiding together (green, violet and brown for $3\beta$, $4\beta$ and $5\beta$ respectively). We notice that for CB and STR phases more than $60\%$ of the configurations are trivial braids while most of the remaining ones only possess one or few permutation cycles of length $2\beta$. In the VBS case, $57\%$ of configurations possess permutation cycles of length $3\beta$ consistent with local resonances harbored in isolated hexagons occupied by three particles~\cite{IsakovPRL2006,LianaPRB}. Finally, permutation cycles of length  $4\beta$ and $5\beta$ only appears in the $\mathbb{Z}_2$ phase with, in some cases, $30-40\%$ of particles involved in permutation cycles longer than 1$\beta$.  Though with the current available tools one is unable to detect braiding between permutation cycles, the observation that longer permutation cycles only appear in the $\mathbb{Z}_2$ phase and a considerable number of particles are involved in them is consistent with the presence of more complicated braids which extend throughout the lattice as discussed in Section~\ref{topo_str}.  Overall, all these observations are consistent with our expectations, as discussed in the previous Section.

\begin{figure}
    \centering
    \includegraphics[width=0.9\textwidth]{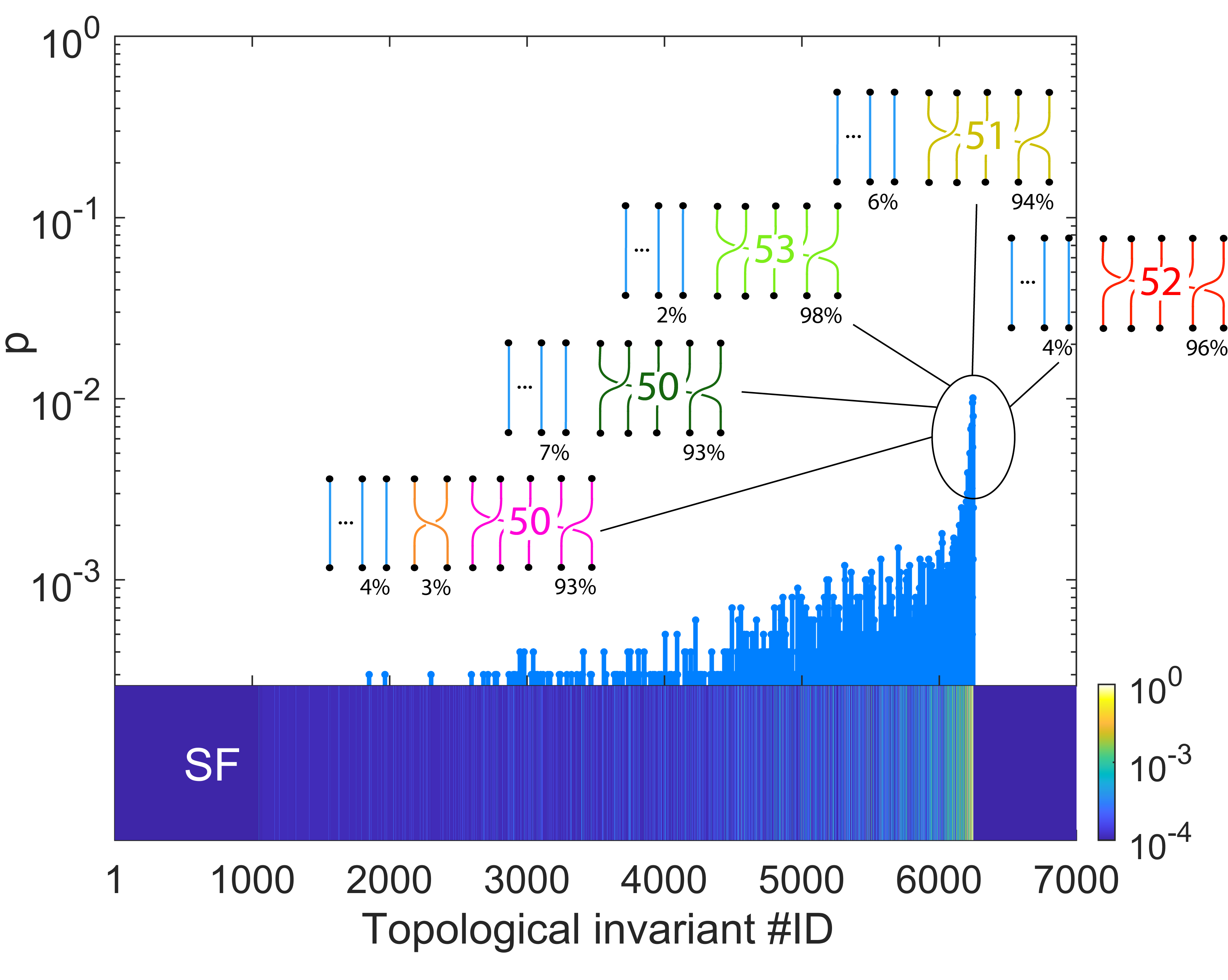}
    \caption{Distribution $p_\mathcal{\tilde T}$ of the ground-state of a superfluid phase (SF) on Kagome lattice ($V/t=5$, $n=1/2$). Top panel shows the five most likely arrangements of permutation cycles.
    Percentages below the geometric braids represent the fraction of particles involved in permutation cycles of that length.}
    \label{fig:SFtopospectrum}
\end{figure}

In Fig.~\ref{fig:SFtopospectrum}, we plot the spectrum of the superfluid phase (SF) ground-state. As expected, we notice that configurations featuring permutation cycle lengths
$\lambda/\beta \sim N$ are the most probable, that is, configurations are such that most worldlines participate to the same permutation cycle. In Fig.~\ref{fig:SFtopospectrum}, we also sketch some of the most probable permutation cycle lengths.
The percentages in the sketches represent the fraction of particles involved in the permutation cycles.

\begin{figure}
    \centering
    \includegraphics[width=0.9\textwidth]{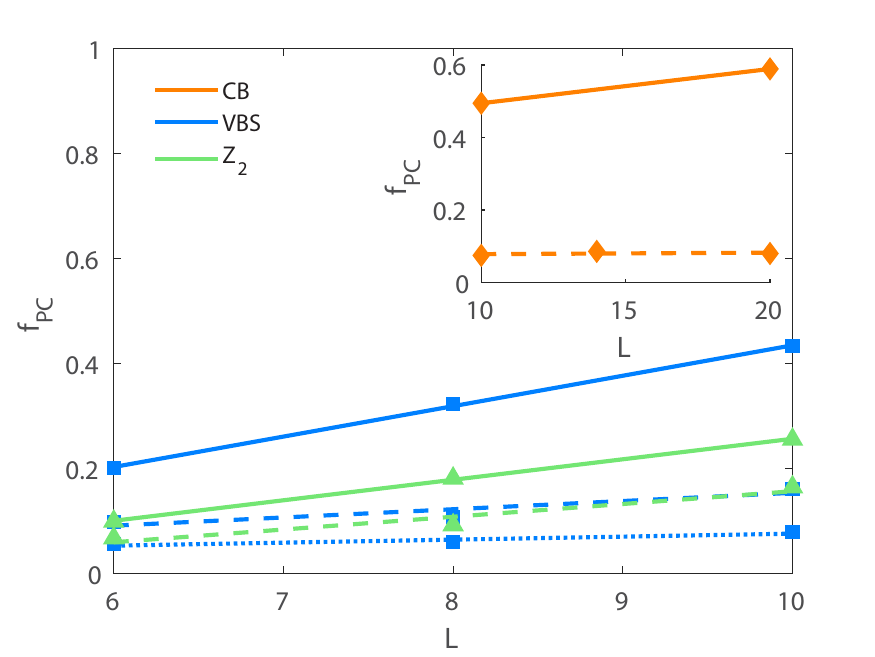}
    \caption{Fraction of particles in permutation cycles longer than $1\beta$ as a function of system size $L$ for CB, VBS and $\mathbb{Z}_2$ ground states.
    CB at $V/V_c=1.7$ (continuous) and $V/V_c=10$  (dashed); VBS at $V/V_c=2$ (continuous), $V/V_c=3$ (dashed) and $V/V_c=4$ (dotted); $\mathbb{Z}_2$ at $V/V_c=2$ (continuous) and $V/V_c=3$ (dashed). }
    \label{fig:fracPC}
\end{figure}
We also looked at how the complexity of topological structures changes with system size. We quantified the complexity by measuring $f_{PC}$ the fraction of particles in permutation cycles longer than $1\beta$ (see Fig.~\ref{fig:fracPC}). We found that deep in the CB and STR phases ($V/V_c=10$ and $V/V_c\approx4$ respectively), $f_{PC}$ does not change as L is increased. This is not surprising, since deep in these solid phases, one does not expect complex braids to appear, as discussed in Section~\ref{topo_str}. This is no longer true as one gets closer to the transition to the SF phase. Here, the complexity of the topological structure increases with L. Larger L implies a larger Hilbert space which, in general, leads to a considerable more complex ground-state expansion. A similar observation is valid for the VBS. While  $f_{PC}$ increases with L at $V/V_c=2$ (solid blue line), this trend becomes less pronounced as   $V/V_c$ is increased and at $V/V_c=4$ (dotted blue line) the L-dependence of $f_{PC}$ has become minimal. The situation for the $\mathbb{Z}_2$ phase at 1/3 filling seems more complex. At $V/V_c=2$ (solid green line), we observe a strong L-dependence of $f_{PC}$. As $V/V_c$ is increased to 3 (dashed green line), the L-dependence  remains relatively pronounced and certainly more pronounced than in the VBS case at the same value of $V/V_c$.
This observation along with the observed arrangement of permutation cycles discussed in Table of Fig.~\ref{fig:topostruct} seems to indicate that the $\mathbb{Z}_2$ ground-state retains more complex braids deeper in the phase, as one would expect for this  spin liquid phase.

\section*{Conclusive remarks}
In this manuscript, we have studied the statistics of a simple topological invariant (in terms of permutation cycles) associated to the braiding structure of worldline configurations for a variety of many-body ground-states. Configurations are collections of particle paths and  result from the path-integral formulation of statistical mechanics.  We have considered hard-core bosons for which configurations can be seen as geometric braids and therefore be assigned a topological structure.
By means of Monte Carlo calculations, we have studied checkerboard, stripe, valence-bond solids, $\mathbb{Z}_2$ topologically ordered spin liquid, and superfluid phase and observed that distinct topological structures correspond to ground-states of different quantum phases.

In the future, we plan to extend this work and further explore the possibility that mathematical properties  of  worldline configurations  are implicitly linked to essential features of quantum states as we believe that pursuing this perspective may provide new insights on phases of matter of quantum many-body systems.

More specifically, we would like to formally show that worldline configurations can be interpreted as visualizations of generation of quantum fluctuations. For the systems considered, quantum fluctuations can be understood in terms of fluctuations of the local occupation number. Quantum fluctuations specific to a certain many-body system collectively determine the macroscopic behavior of the system as described by its quantum order or, equivalently, its entanglement~\cite{anyon1,kitaev2006,wen2010,wen2019}. We believe the proposed interpretation may provide a new scenario for the characterization of certain aspects of ground-state entanglement of hardcore lattice bosons and may also inspire how to think of many-body entanglement for a broader class of models~\cite{Sirker_2012}.
In Schr\"odinger's own words, quantum entanglement ``is not one but rather \emph{the} characteristic trait of quantum mechanics''.
In order to acquire knowledge on how the information and correlations are organized among the constituents of a quantum state, it is essential to study the way the state is entangled. We plan to further develop our method in order to more completely characterize the topological structure and  establish a more explicit connection between relevant sets of configurations and their topological structures with the inseparability of the quantum state, i.e. to its entanglement. The high classificability of configurations in terms of their topological and geometrical properties has the potential to go beyond the mere differentiation of ground-states, and to engineer beyond-scalar and more complex quantities capable to better address and represent the vast complexity of  entangled states.

\bigskip
\noindent
\textbf{Data Availability.} The datasets generated during the current study are available from the corresponding author on reasonable request.

\section*{Acknowledgements}
The authors wish to thank Vittorio Penna for fruitful discussions.
W.W. thanks Anne E. B. Nielsen and Yi Li for motivation on considering topological properties of worldline configurations. The computing for this project was performed with the High Performance Computing Cluster at Clark University.

\section*{Author contributions statement}
B.C.S., F.L, and W.W. conceived the main ideas, discussed the results, and edited and reviewed the manuscript. W.W. provided the mathematical intuition. F.L. performed the investigation and the data analysis. L.S. developed the code to extract statistics of permutation cycles from MC configurations. B.C.S. supervised the entire project. F.L and W.W. equally contributed to the work.

\section*{Additional Information}
\textbf{Competing Interests:} The authors declare that they have no competing interests.

\appendix
\section{Definition of topological structure}\label{App_A}
We characterize the topological structure of geometric braids by homotopy on them rather than isotopy (on which the standard definition of the braid group is based). While in both cases topologically-equivalent geometric braids can be continuously deformed into each other, isotopy further requires the ``ends'' (i.e state $\ket\alpha$ in configurations) of topologically equivalent geometric braids to be fixed. Here, we do not need this additional constraint because, due to periodic boundary conditions in imaginary time, $\ket\alpha$ has no special meaning in the sequence of states specifying a certain configuration. In other words, any configuration corresponding to a cyclic permutation of a given sequence of states has exactly the same weight. Moreover, configurations which only differ by spatial position of braiding events have an equal weight due to lattice symmetry.
\section{Properties of configurations}\label{AppA0}
Ground-state expansion coefficients can be expressed using Eq.~\ref{eq4} and replacing operator $O$ by $|\alpha\rangle\langle\alpha|$:
\begin{equation}
|c_\alpha|^2=\expval{\Psi\dyad{\alpha}\Psi}=\lim_{\beta\rightarrow\infty}\mel{\alpha}{\frac{\mathrm{e}^{-\beta H}}{\mathcal{Z}_\beta}}{\alpha}=\lim_{\beta\rightarrow\infty}\int_{\phi\in\mathcal{C}} ~ \frac{\omega_\phi \dyad{\alpha}{\alpha}_\phi}{\mathcal{Z}_\beta},\label{calpha}
\end{equation}
where $\dyad{\alpha}{\alpha}_\phi=1$ when $\phi$ starts with $\ket{\alpha}$, and $0$ otherwise.
The actual computation of the set of $c_\alpha$ is in general a numerical intractable as the Hilbert space is too large. Nonetheless, we can make some simple considerations: (i) Expansion coefficients are positive-definite since $\omega_\phi$ are ($t_{ij}>0$ for the considered models); (ii) if a state $\ket{\alpha^\prime}$ does not enter the expansion of the ground-state ($c_\alpha^\prime=0$), all the configurations starting and ending with $\ket{\alpha^\prime}$ would have weight $\omega_\phi\rightarrow0$. In calculating the partition function $\mathcal{Z}_\beta$, periodic boundary conditions in imaginary time hold, this means that any configurations containing $\ket{\alpha^\prime}$ in the middle of the sequence would also have $\omega_\phi\rightarrow0$. For this reason, configurations $\phi$ may  be  relevant  to  the ground-state expansion only if all $\ket{\alpha_m}$ in $\phi$ contribute to the expansion of $\ket{\Psi}$.

\section{Labelling of topological structures}\label{App_C}

In general, configurations belonging to different ground-states correspond to geometric braids that cannot be compared with each other as the number $N$ of particles required to stabilize these phases may be different from case to case. In other words, they belong to different braiding groups $\mathcal{B}_N(\mathcal{S})$.
However, since in insulating ground-states the $1\beta$-long permutation cycle (PC) is always macroscopically occupied, one may only consider the geometric braids between the $N^\prime$ particles that do not participate in the $1\beta$-long PC. This can be done by first redefining topological invariant $\vec {q}_\phi$  as
\begin{equation}
    \vec{p}_\phi=(..., n_l\times l ...)/N \label{pfi} \; ,
\end{equation}
so that component $l$ of ${\vec p_\phi}$ is the fraction of particles (worldlines) involved in  permutation cycles of length $l\time\beta$. Next, we introduce a new topological invariant $\vec{p}_\phi^{\;\prime}=(2\times n_2,3\times n_3 \dots N\times n_N)/N$ (note that the 1$\beta$ long PC is not included) corresponding to geometric braids belonging to $\mathcal{B}_{N^\prime}(\mathcal{S})$. By only considering the fraction of particles involved in PC longer than 1$\beta$, it is possible to better compare the arrangement of permutation cycles in different ground-states.

Finally, we introduce a topological invariant  which can be interpreted as an `average' length of PC useful to order and label different $\vec{p}_\phi$:
\begin{equation}
    \langle\lambda_\phi\rangle=\vec{p}_\phi^{\;\prime}\cdot\vec{\lambda} \label{avgL}
\end{equation}
where $\vec{\lambda}=(2, ..., N)$ is the vector of PC lengths and symbol $\cdot$ refers to dot product.
As an example, configurations $\phi_1=(0,0,1)$ and $\phi_2=(1,1,0)$ of Fig.~\ref{fig:1} ($N=3$) will feature $\vec{p}_{\phi_1}^{\;\prime}=(0,1)$ and $\vec{p}_{\phi_2}^{\;\prime}=(2/3,0)$, resulting in
\begin{eqnarray}
    &&\langle\lambda_{\phi_1}\rangle=(0,1)\cdot(2,3)=3,\\
    &&\langle\lambda_{\phi_2}\rangle=(2/3,0)\cdot(2,3)\approx1.33.
\end{eqnarray}
A generic configuration $\chi$ with $100\%$ $1\beta$-long permutation cycles will have $\vec{p}_{\chi}^{\;\prime}=\vec{0}_{N-1}^{\,}$ (i.e. the $N-1$ dimensional null vector) and $\langle\lambda_{\chi}\rangle=0$.
The integer ID labels organizing the topological spectra shown in Fig. 2 and Fig. 4 result from ordering configurations according to (\ref{avgL}).

\section{$V_c/t$ values} \label{AppE}
We report in table~\ref{tab:Vc} the critical values $V_c/t$ at which the transitions to the phases studied in the paper occur.
\begin{table}[h!]
    \centering
    \begin{tabular}{ c|c }
        Phase Transition & $V_c/t$  \\
        \hline
        SF to CB & 2.00 ~\cite{Hebert2001}\\
        SF to STR & 12.5~\cite{Capogrosso2010}\\
        SF to VBS & 7.796~\cite{IsakovPRL2006}\\
        SF to $\mathbb{Z}_2$ & 7.0665~\cite{Isakov2011,Roychowdhury2015}\\
    \end{tabular}
    \caption{Critical values $V_c/t$ for the considered phases. For uncertainties and further details we remand to the references. Notice that the $V_c/t$ from~\cite{Capogrosso2010} refers to simulations with no cutoff in the range of the interaction. Here, we put a cutoff to four lattice spaces, so the exact position of the transition will be slightly different. }
    \label{tab:Vc}
\end{table}


\bibliography{Refs.bib}

\end{CJK*} 
\end{document}